\pdfoutput=1

  \documentclass[aps, prd, twocolumn, nofootinbib, longbibliography]{revtex4-1}
 \newcommand{\aap}{Astronomy and Astrophysics}
 \newcommand{\mnras}{Monthly Notices of the Royal Astronomical Society}
 \newcommand{\jcap}{Journal of Cosmology and Astroparticle Physics}

\usepackage{natbib}
\usepackage{amsmath}
\usepackage{float}
\usepackage[colorlinks=true, allcolors=blue]{hyperref}
\usepackage{graphicx}
\usepackage[english]{babel}

\newcommand{\be}{\begin{equation}}
\newcommand{\ee}{\end{equation}}
\newcommand{\ba}{\begin{eqnarray}}
\newcommand{\ea}{\end{eqnarray}}

\newcommand{\LCDM}{$\Lambda$CDM }

\begin{document}

\title{Data compression in cosmology: 
A compressed likelihood for \textit{Planck} data}

\author{Heather Prince}
\email{heatherp@princeton.edu}
\author{Jo Dunkley}
\affiliation{Department of Astrophysical Sciences, Peyton Hall,
Princeton University, Princeton, New Jersey 08544, USA }

\begin{abstract}

We apply the Massively Optimized Parameter Estimation and Data compression
technique (MOPED) to the public \textit{Planck} 2015 temperature likelihood,
reducing the dimensions of the data space to one number per parameter of
interest. We present CosMOPED, a lightweight and convenient compressed
likelihood code implemented in Python.
In doing so we show that the $\ell<30$ \textit{Planck} temperature likelihood
can be well approximated by two Gaussian distributed data points,  which allows
us to replace the map-based low-$\ell$ temperature likelihood by a simple
Gaussian likelihood. We make available a Python implementation of
\textit{Planck}'s 2015 Plik\_lite temperature likelihood that includes these
low-$\ell$ binned temperature data (Planck-lite-py). We do not explicitly use the
large-scale polarization data in CosMOPED, instead imposing a prior on the
optical depth to reionization derived from these data. We show that the \LCDM
parameters recovered with CosMOPED are consistent with the uncompressed
likelihood to within 0.1$\sigma$, and test that a 7-parameter extended model
performs similarly well.

\end{abstract}

\maketitle

\section{Introduction}
When analyzing a large dataset with many different data points, but only a few
relevant parameters, it can be useful to compress the data, retaining all the
information that is relevant to the parameters of interest. These compressed
data can then be used for parameter inference.
\citet{tegmark_karhunen-loeve_1997} showed that if there is one parameter of
interest, a dataset can be compressed to one number without losing any
information about that parameter. The Massively Optimized Parameter Estimation
and Data compression technique (MOPED) described in
\citet{heavens_massive_2000} extends this to multiple parameters, developing
optimal linear compression for Gaussian data provided the covariance is
independent of the parameters. \citet{alsing_generalized_2018} show that
compression to the score function (the gradient of the log likelihood)
preserves the information content even for non-Gaussian data and data for which
the covariance depends on the parameters. Thus the full dataset of $N$ data
points can be compressed to $n$ numbers, where $n$ is the number of parameters of
interest, while preserving the Fisher information about those parameters.

This compression step is useful both as a way to generate a simplified
likelihood function, and as a step towards likelihood-free inference when the
form of the likelihood is not known precisely
\citep[e.g.,][]{alsing_generalized_2018}. Compressions of this form have been
used to study the star formation history of galaxies
\citep{reichardt_recovering_2001, heavens_star-formation_2004, panter_star_2007}
 and considered for exoplanet transit detections \citep{protopapas_fast_2005},
 gravitational wave studies with {\it LISA} \citep{graff_investigation_2011},
 and covariance matrix estimation \citep{heavens_massive_2017}.
 \citet{gupta_fast_2002} proposed applying MOPED compression to cosmic
 microwave background (CMB) data, and \citet{zablocki_extreme_2016} applied a
 similar compression scheme to the temperature power spectrum of {\it WMAP}.

In this paper we apply the MOPED compression to the \textit{Planck} CMB power
spectrum \cite{planck2015_like}, and show that the standard \LCDM\ cosmological
parameter constraints can be derived from a compressed likelihood of just six
Gaussian-distributed data points. We go beyond earlier analyses of CMB data by
precompressing the non-Gaussian large-scale temperature power spectrum into
two approximately Gaussian data points. We make the software for the
MOPED-compressed likelihood publicly available, as well as for the likelihood
computed directly from the binned power spectrum with the inclusion of new
large angular scale bins.
\footnote{At \href{https://github.com/heatherprince/cosmoped}
{https://github.com/heatherprince/cosmoped} and
\href{https://github.com/heatherprince/planck-lite-py}
{https://github.com/heatherprince/planck-lite-py}.
Both codes have been updated with the 2018 temperature and polarization Plik-lite
likelihood}

The outline of the paper is as follows. In \S\ref{sec:planck} we describe the
public \textit{Planck} likelihood and the precompression we implement to
better approximate it as Gaussian. In \S \ref{sec:moped} we describe the MOPED
compression scheme we apply, and in \S \ref{sec:results} show parameter
constraints for the \LCDM\ model and an example extended model with running of
the primordial spectral index. We conclude in \S \ref{sec:conclude}.

\section{\textit{Planck} likelihood and low-$\ell$ binning}
\label{sec:planck}

\begin{figure}[!htp]
  \centering
  \includegraphics[width=\linewidth]{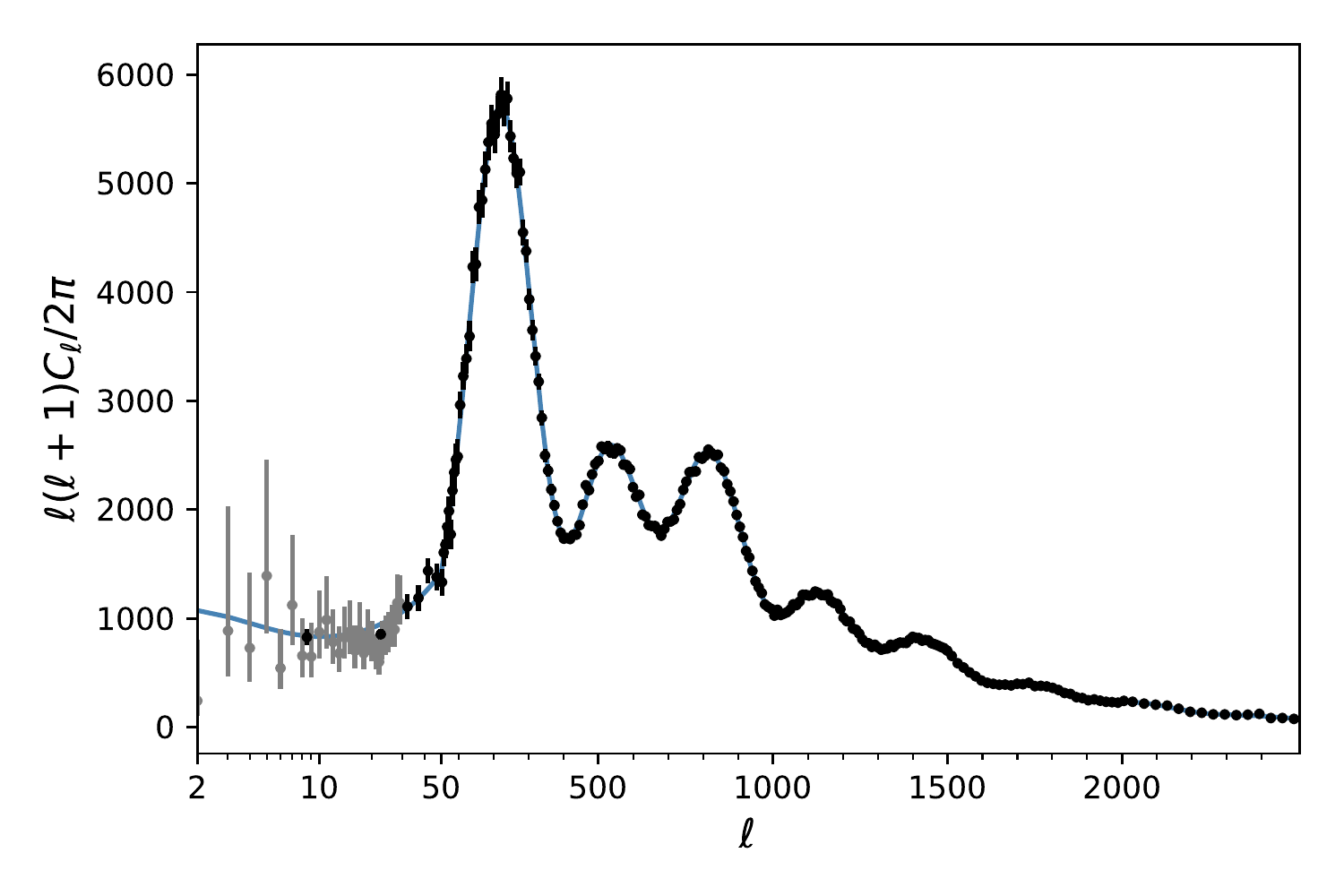}
  \caption{The \textit{Planck} 2015 temperature power spectrum from
  \protect{\citep{planck2015_like}}. The non-Gaussian $\ell<30$ bandpowers are shown with
  their asymmetrical errors in gray.
  The two black points at $\ell<30$ are obtained by estimating the binned
  spectrum in two angular bins, and have approximately Gaussian distributions.
  We use these two low-$\ell$ bins for the likelihoods described in this paper.
  The $\ell \geq 30$ binned data are the foreground-marginalized temperature
  bandpowers from the Plik\_lite likelihood for \textit{Planck} 2015. The
  theoretical power spectrum for the \textit{Planck} 2015 TT+lowTEB best fit
  parameters \protect{\citep{planck2015_parameters}} computed using CLASS \protect{\citep{CLASS2011}} is shown in blue.}
   \label{fig:spectra}
\end{figure}

The current state-of-the-art CMB data come from the \textit{Planck} satellite.
The latest cosmological analysis is reported in \cite{planck2018_cosmo}, with
public data from the earlier 2015 analysis described in \cite{planck2015_like}.
The \textit{Planck} temperature power spectrum is shown in
Fig. \ref{fig:spectra}.
The likelihood function, describing the probability of the data given some
model, is separated into two main parts for the \textit{Planck} power spectrum
analysis, with different approaches for large scales and smaller scales. We
summarize these components briefly here.

At $\ell \geq 30$ (corresponding to scales smaller than several degrees on the
sky) the likelihood $\mathcal{L}$ for the temperature and $E$-mode polarization
power spectra (TT, TE, and EE) is modeled as a Gaussian distribution, with
\be
-2 \ln \mathcal{L} = (C_b^{\rm th} - C_b^{\rm data})^T Q^{-1}(C_b^{\rm th} -
C_b^{\rm data})
\ee
to within an overall additive constant, with binned data $C_b^{\rm data}$,
binned theory $C_b^{\rm th}$, and binned covariance matrix $Q$. For the
Plik\_lite likelihood \citep{planck2015_like}, these data spectra represent an
estimate of the CMB bandpowers, with foregrounds already marginalized over
using the approach of \citet{Dunkley2013}.

At $\ell<30$ the distribution of the angular power spectrum is non-Gaussian.
For the 2015 data release, the \textit{Planck} team released a joint pixel
based likelihood for temperature and polarization for $\ell \leq 29$ (`lowTEB').
 There is also a standalone temperature low-$\ell$ likelihood based on the
foreground-cleaned Commander temperature map, which we use in this paper.
These likelihoods are computed in map space since the distribution of the
power spectrum on these scales is non-Gaussian. The 2018 likelihood uses a
similar low-$\ell$ temperature likelihood, and a separate low-$\ell$
polarization likelihood built from simulations \citep{planck2018_like}.

\subsection{Low-$\ell$ temperature bins}
The compression approach we adopt, which we describe in the next section, is
optimal for Gaussian distributions. Since we are interested in a lightweight
compression to estimate simple cosmological models, we first compress the
$\ell<30$ \textit{Planck} TT data into two bins with approximately Gaussian
distributions. We do this by conditionally sampling the posterior distribution
for the power in each bin, estimating
\be
p(\theta|d) \propto p(d|\theta)p(\theta).
\ee
Here the parameters $\theta$ are the binned values $D_{2\leq \ell\leq 15}$ and
$D_{16\leq \ell\leq 29}$, where $D_\ell = \ell (\ell+1) C_\ell /2 \pi$,
assuming a constant value for $D$ in each bin. The likelihood $p(d|\theta)$ is
the \textit{Planck} $\ell<30$ temperature likelihood function. We assume
uniform priors on $\theta$. The binning is performed on $D_\ell$ rather than
$C_\ell$ because $D_\ell$ is approximately constant for the low-$\ell$
temperature power spectrum. When binned values of $C$ are required we convert
from the binned $D$ values by dividing by the mean of $\ell(\ell+1)/2\pi$ in
the bin.

\begin{figure}[!th]
  \centering
  \includegraphics[width=\linewidth]{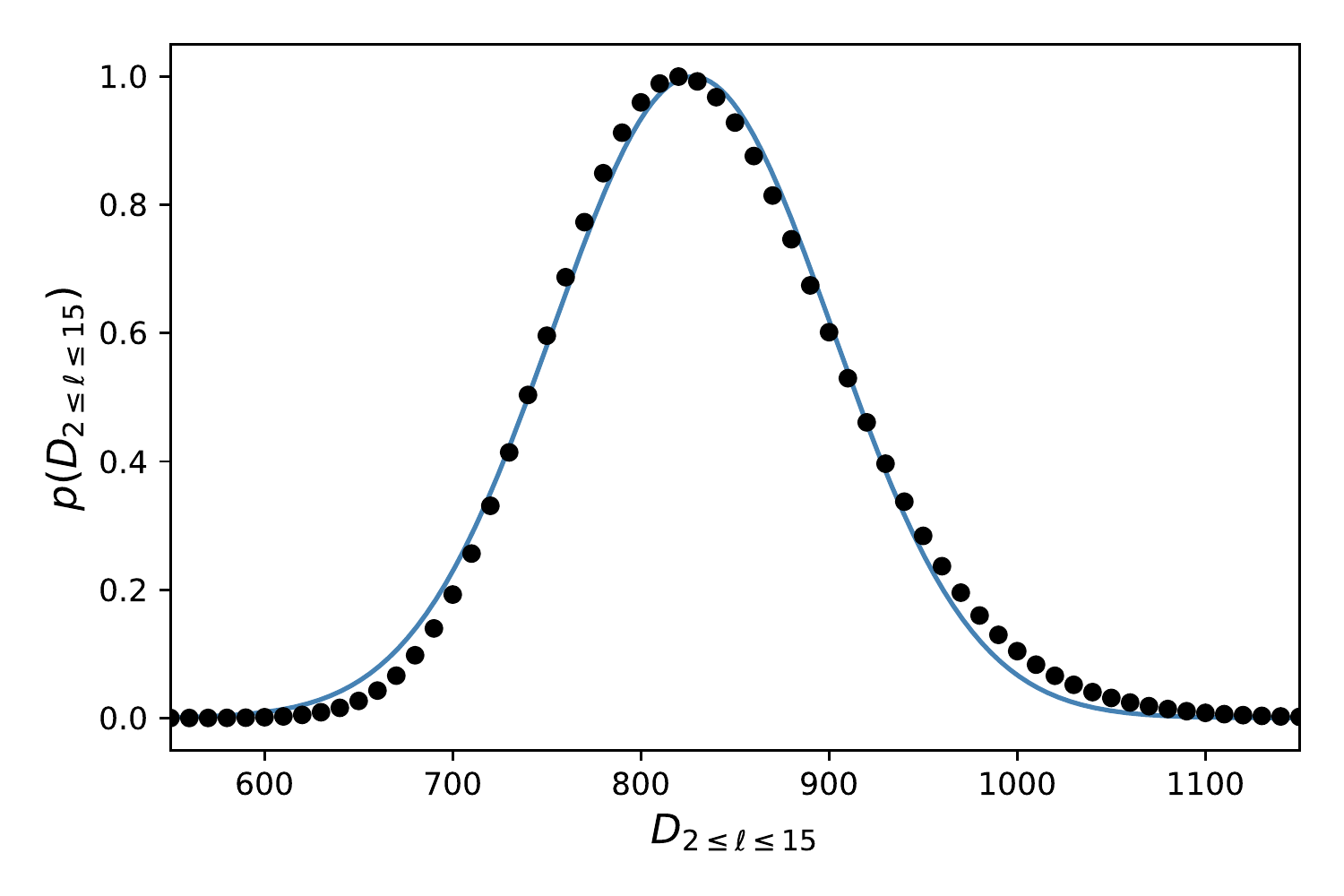}
  \includegraphics[width=\linewidth]{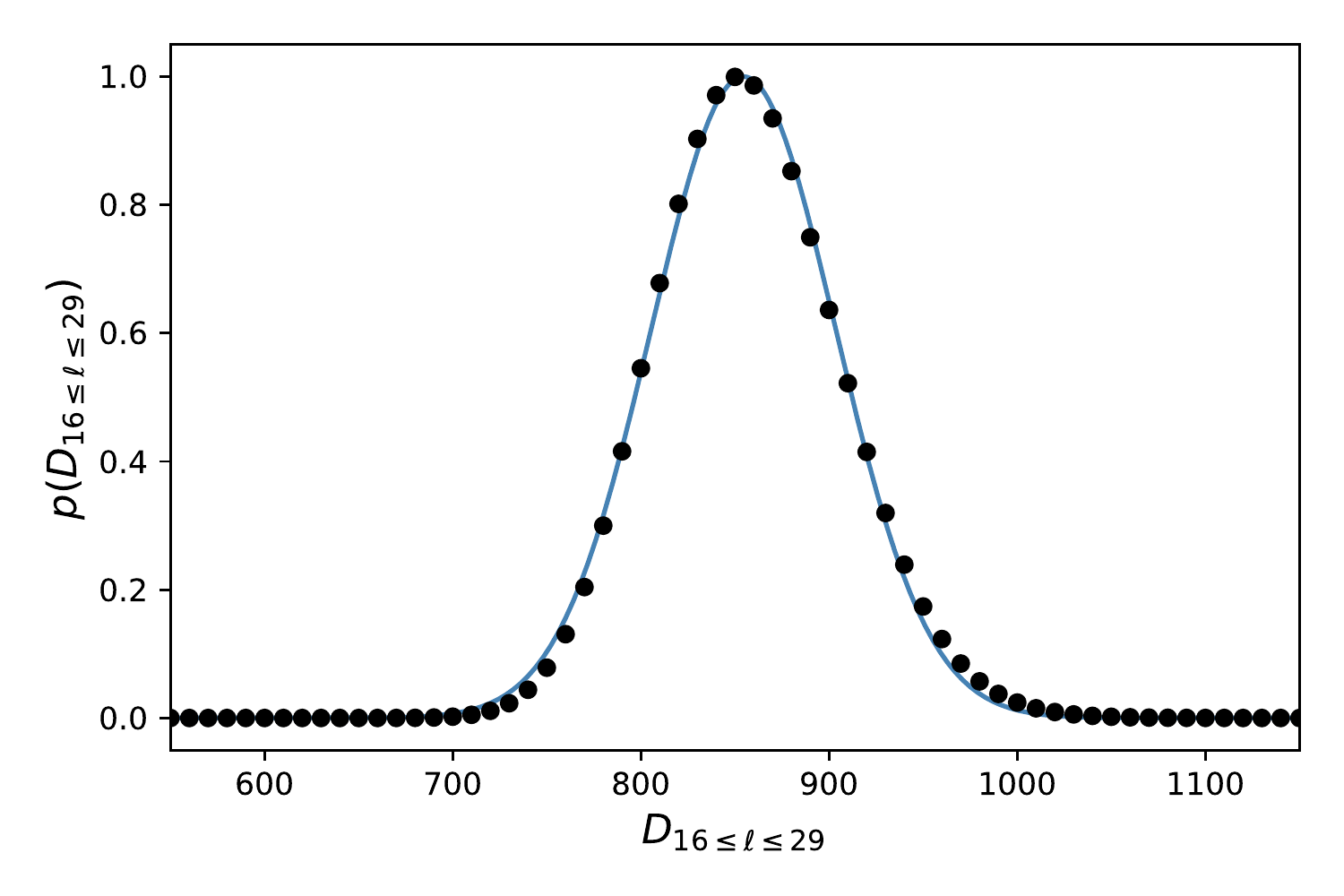}
  \caption{The probability distribution for $D_\ell$ for two low-$\ell$
  temperature bins using the \textit{Planck} 2015 low-$\ell$ Commander
  likelihood. The best-fit Gaussian distribution is shown in blue.}
   \label{fig:lowell}
\end{figure}

The distributions of the two low-$\ell$ power spectrum bins are shown in Fig.
\ref{fig:lowell}, together with the best-fitting Gaussian distributions.
We find
\ba
D_{2\leq \ell\leq 15} = 827\pm74~\mu {\rm K}^2\nonumber\\
D_{16\leq \ell\leq 29} = 854\pm49~\mu {\rm K}^2.
\ea
Their distributions are close to Gaussian, unlike the distributions for the
individual multipoles. These bandpowers are also indicated in
Fig. \ref{fig:spectra} (the first two black points), together with the
unbinned low-$\ell$ power spectrum in gray.
We chose this binning scheme before sampling parameters; other choices that 
produce approximately Gaussian distributions would be expected to give similar
results.

\subsection{Low-$\ell$ polarization ($\tau$ prior)}
The amplitude of the large-scale polarization signal depends primarily on the
optical depth to reionization, as well as the primordial amplitude. To include
the low-$\ell$ polarization data we compress the 2015 polarization information
into a single Gaussian prior on the optical depth to reionization, adopting
$\tau= 0.067 \pm 0.023 $ derived from the \textit{Planck} low-$\ell$ likelihood
using the Low Frequency Instrument \citep{planck2015_like}. This is an
approximation since $\tau$ is correlated with other cosmological parameters,
in particular the primordial amplitude $A_s$.
Improved measurements of the optical depth have since been made from the
\textit{Planck} High Frequency Instrument
\cite{planck2016_hfi, planck2018_cosmo}.
However, the purpose of this study is to compress the public 2015 likelihood,
and we defer a future refinement of our compression code to include the 2018
polarization information that was recently made public.

\subsection{Parameter constraints}
\begin{figure*}[!htb]
  \centering
  \includegraphics[width=\linewidth]{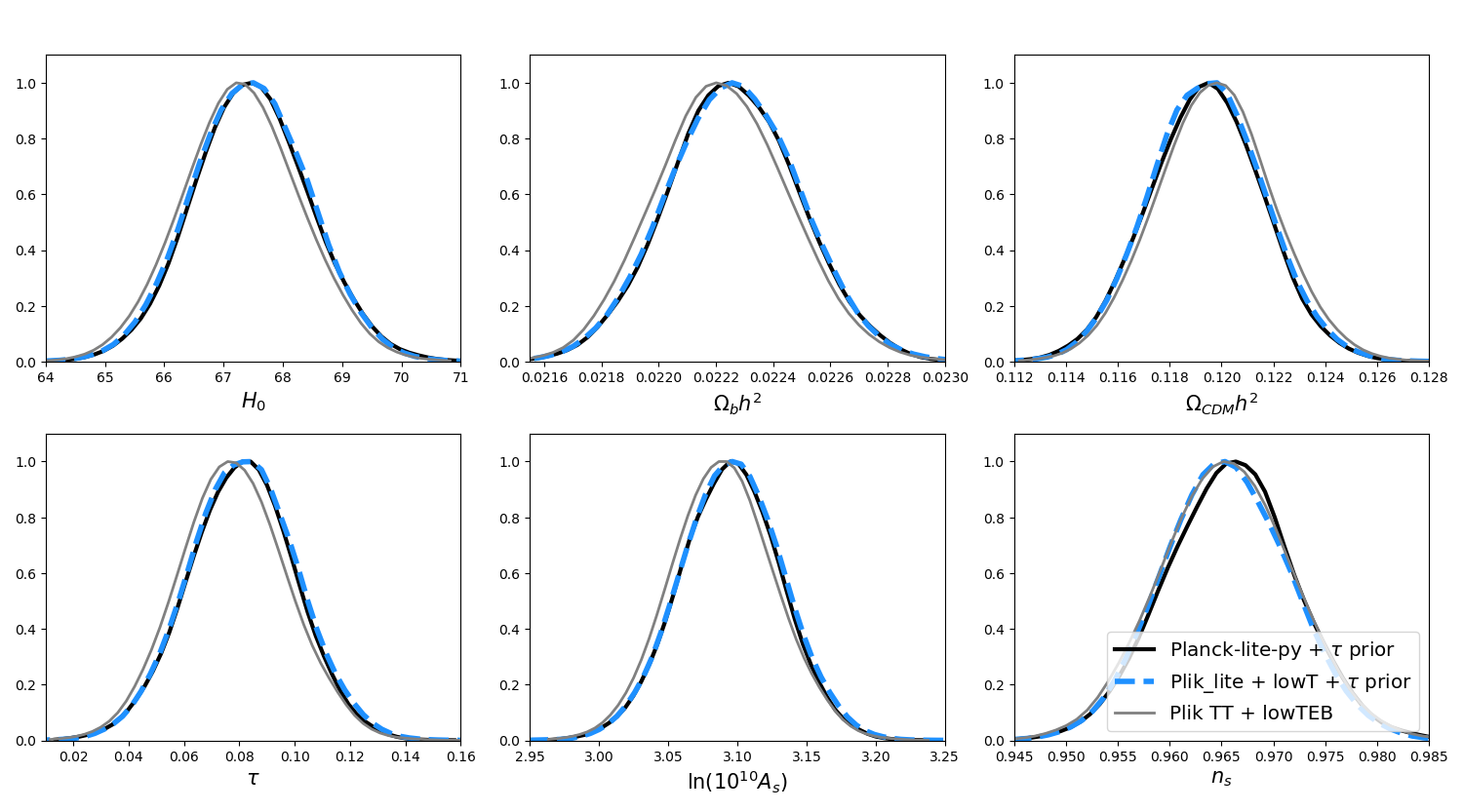}
  \caption{Posteriors on the \LCDM parameters from Planck-lite-py, a Python
  implementation of the \textit{Planck} Plik\_lite likelihood with two Gaussian
  low-$\ell$ bins (black solid curve) compared to the combined Plik\_lite TT +
  low-$\ell$ Commander \textit{Planck} 2015 temperature likelihoods
  (blue-dashed curve). The parameter constraints agree to within 0.1$\sigma$,
  showing that our two binned low-$\ell$ data points capture enough information
  to give equivalent constraints to the full low-$\ell$ temperature likelihood.
  A Gaussian prior of $\tau=0.067 \pm 0.023$ is included in the black and blue
  curves. We also show results from  the public chains from sampling the
  TT+lowTEB \textit{Planck} 2015 likelihood (gray solid curve) for reference.
  The small difference is the effect of imposing a prior on the optical depth.}
   \label{fig:params_two_low_ell_bins}
\end{figure*}

The effect of describing the low-$\ell$ temperature data using two Gaussian
bins and using a prior on $\tau$ in place of the low-$\ell$ polarization likelihood is shown in Fig. \ref{fig:params_two_low_ell_bins}, which shows the
posterior probabilities for the six \LCDM  parameters (the Hubble constant,
baryon density, cold dark matter density, amplitude and spectral index of
primordial fluctuations, and optical depth to reionization) obtained by
sampling three different likelihood combinations.

Replacing the low-$\ell$ temperature likelihood with a Gaussian likelihood
based on two low-$\ell$ temperature bins results in parameter constraints
that agree well, to within 0.1$\sigma$ (black versus blue-dashed in
Fig. \ref{fig:params_two_low_ell_bins}).
Here, the black solid curve is derived using our Python implementation of
Plik\_lite with the additional two Gaussian low-$\ell$ bins included
(Planck-lite-py).  The blue-dashed curve shows the posteriors obtained by
sampling the \textit{Planck} high-$\ell$ temperature Plik\_lite and low-$\ell$
temperature-only Commander likelihoods using the CosmoSIS cosmological
parameter estimation code \citep{Zuntz2015}. In both cases we compute the
theoretical CMB power spectrum using the Cosmic Linear Anisotropy Solving
System (CLASS) \citep{CLASS2011}, sample the likelihood using the emcee
\citep{Foreman-Mackey2013} Python implementation of Markov chain Monte Carlo
(MCMC) affine-invariant ensemble sampling \citep{GoodmanWeare2010}, and impose
a Gaussian prior on the optical depth $\tau= 0.067 \pm 0.023 $ from the
\textit{Planck} low-$\ell$ likelihood \citep{planck2015_like}. The low-$\ell$
data provide an important anchor for constraints on parameters such as spectral
index $n_s$ and Hubble parameter $h$. The Planck-lite-py parameter constraints
agree with the Plik\_lite+lowT constraints to within 0.1$\sigma$, showing that
for the standard cosmological model the full low-$\ell$ temperature likelihood can be
replaced by a Gaussian likelihood with two bins and a diagonal covariance
matrix without a significant effect on the parameter constraints.

The gray curve in Fig. \ref{fig:params_two_low_ell_bins} shows the public
\textit{Planck} 2015 chains from the TT+lowTEB likelihood. When comparing
parameters constraints from the full TT+lowTEB likelihood with the
TT+lowT+$\tau$ prior likelihood some of the parameters shift by up to
0.2$\sigma$ because we are using a tau prior which is only an approximation
to the full low-$\ell$ polarization likelihood.

We make available Planck-lite-py, a Python implementation of the \textit{Planck}
Plik\_lite likelihood that includes these low-$\ell$ binned temperature data.
\footnote{\href{https://github.com/heatherprince/planck-lite-py}
{https://github.com/heatherprince/planck-lite-py}.}

\section{MOPED compression vectors}
\label{sec:moped}

We use the Massively Optimized Parameter Estimation and Data compression
technique (MOPED) described in \citet{heavens_massive_2000} to compress the
\textit{Planck} power spectrum. This linear compression is optimal in the sense
that the Fisher information content is preserved for Gaussian data when the
covariance matrix is independent of the parameters of interest.

The \textit{Planck} data vector used in this analysis is the
foreground-marginalized binned temperature angular power spectrum. It would be
straightforward to include the TE and EE angular power spectra in the data
vector for a combined temperature and polarization compressed likelihood. The
binning is performed using a constant weighting in $D_\ell$
\citep{planck2015_like}, which corresponds to
\be
C_{b} = \sum\limits_{\ell=\ell_b^{min}}^{\ell_b^{max}} w_b^{\ell} C_\ell,
\label{eq:binning}
\ee
where
\be
 w_b^{\ell} =\frac{\ell(\ell+1)}{\sum\limits_{\ell=\ell_b^{min}}^{\ell_b^{max}}
 \ell(\ell+1)} .
\ee
The binned angular power spectrum is the sum of a signal component that depends
on the cosmological parameters $\boldsymbol{\mu}=C_{b}^{th}$ (the noise-free
theoretical binned angular power spectrum), as well as a noise component
$\boldsymbol{n}$. The total data vector is thus
\be
\boldsymbol{x}=\boldsymbol{\mu}+\boldsymbol{n}.
\ee

The data vector can be compressed into a single number while preserving the
information about the first cosmological parameter of interest $\theta_1$
\cite{tegmark_karhunen-loeve_1997}
\be
y_1=\boldsymbol{b_1}^t\boldsymbol{x}
\ee
with
\be
\boldsymbol{b_1}=\frac{\boldsymbol{Q}^{-1}\boldsymbol{\mu}_{,1}}
{\sqrt{\boldsymbol{\mu}^t_{,1} \boldsymbol{Q}^{-1} \boldsymbol{\mu}^t_{,1}}},
\ee
\vspace{5mm}
where $\boldsymbol{\mu}_{,1}$ is the derivative of the signal (the theoretical
binned temperature angular power spectrum) with respect to the first
cosmological parameter, $\boldsymbol{Q}$ is the covariance matrix, and the
normalization of the compression vector has been chosen such that
$\boldsymbol{b}_1^t \boldsymbol{Q} \boldsymbol{b}_1=1$.

Additional compression vectors can be found that produce linear combinations
which capture information about the other cosmological parameters, while being
orthogonal to the other compression vectors so that each linear combination
$y_m$ is uncorrelated with the others \cite{heavens_massive_2000}, giving
\be
\boldsymbol{b}_m=\frac{\boldsymbol{Q}^{-1}\boldsymbol{\mu}_{,m} -
\sum\limits_{q=1}^{m-1}(\boldsymbol{\mu}_{,m}^t
\boldsymbol{b}_q)\boldsymbol{b}_q} {\sqrt{\boldsymbol{\mu}_{,m}
\boldsymbol{Q}^{-1}\boldsymbol{\mu}_{,m} - \sum\limits_{q=1}^{m-1}
(\boldsymbol{\mu}_{,m}^t \boldsymbol{b}_q)^2}}.
\ee

\begin{figure*}[!htp]
  \centering
  \includegraphics[width=\linewidth]{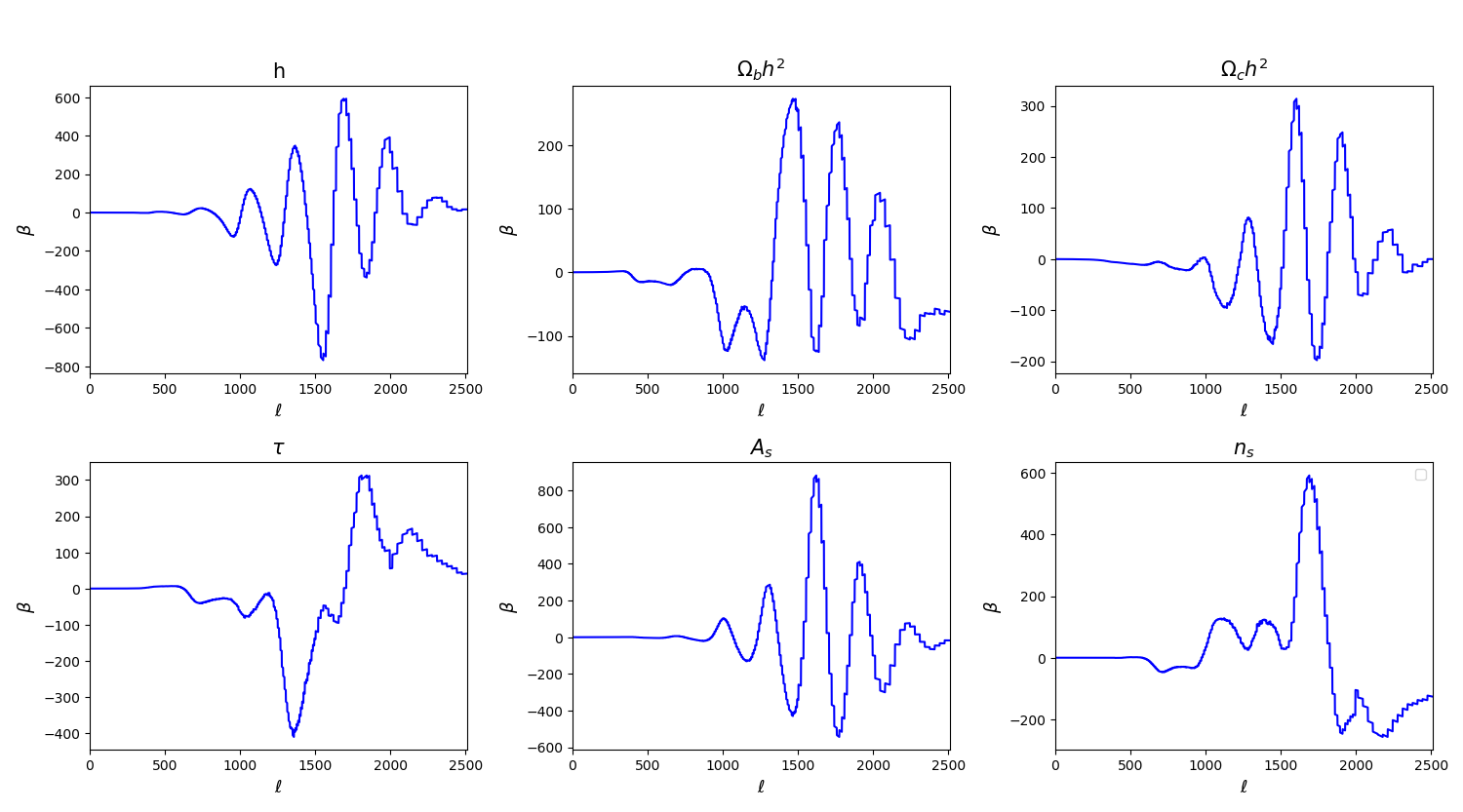}
   \caption{Compression vectors for the \LCDM parameters. The sharp features
   are caused by binning of the power spectrum. These compression vectors can
   be applied to the CMB temperature power spectrum to give six numbers that
   contain as much information about the cosmological parameters as the binned
   temperature power spectrum. The compression vectors depend on the order in
   which they are computed, as they are intentionally orthogonal to one
   another.}
   \label{fig:weight_vectors}
\end{figure*}

\begin{figure}[!htp]
  \centering
  \includegraphics[width=\linewidth]{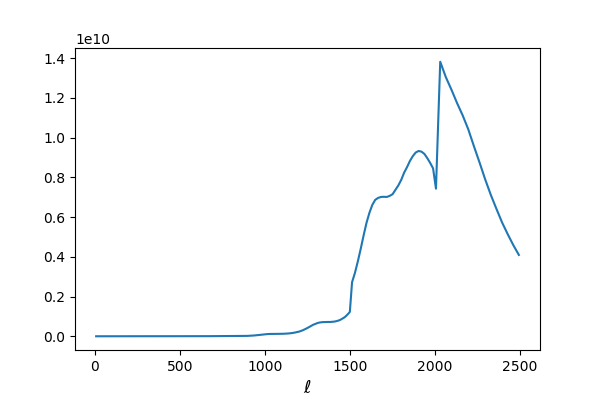}
   \caption{The diagonal of the binned temperature inverse covariance matrix
   for the 217 temperature bins (2 for low-$\ell$ and 215 for high-$\ell$).
   The diagonal elements are small for $\ell<1000$, then rise steeply for
   $1500<\ell<2000$ before dropping again when the experimental noise
   approaches the signal.}
   \label{fig:fisher}
\end{figure}

The compression vectors for the six standard \LCDM parameters are shown in
Fig. \ref{fig:weight_vectors}. The oscillatory behavior comes from the effect
of the acoustic peaks on the derivatives of the theoretical CMB power spectrum.
Most of the signal comes from $1000<\ell<2500$ because this is where the
diagonal of the binned inverse covariance matrix is large
(Fig. \ref{fig:fisher}). At low $\ell$ cosmic variance dominates the noise
while at high $\ell$ experimental noise takes over. The noise in each bin is
also dependent on the bin width, which varies for different multipoles
\citep{planck2015_like}.

Applying these compression vectors to the data vector
\be
y_m=\boldsymbol{b_m}^t\boldsymbol{x}
\ee
gives a set of $M$ numbers $y_m, m=1, ..., M$ which contain as much
information about the cosmological parameters of interest $\theta_m$ as the
full angular power spectrum with $N$ bins. For the binned temperature power
spectrum the data vector has length $N=217$, and for the standard \LCDM
cosmology the number of parameters of interest is $M=6$.

The \textit{Planck} power spectrum and covariance matrix that are used to
compute the compression vectors are already binned, so the compressed
statistics for the data come from applying binned compression vectors to
the power spectrum.
To compress the theoretical CMB power spectrum we use a version of the
compression vectors that includes the binning, weighting each multipole
appropriately as per Eq. (\ref{eq:binning}), so that the binning and
compression are achieved in the same step.

\begin{figure*}[!htp]
  \centering
  \includegraphics[width=\linewidth]{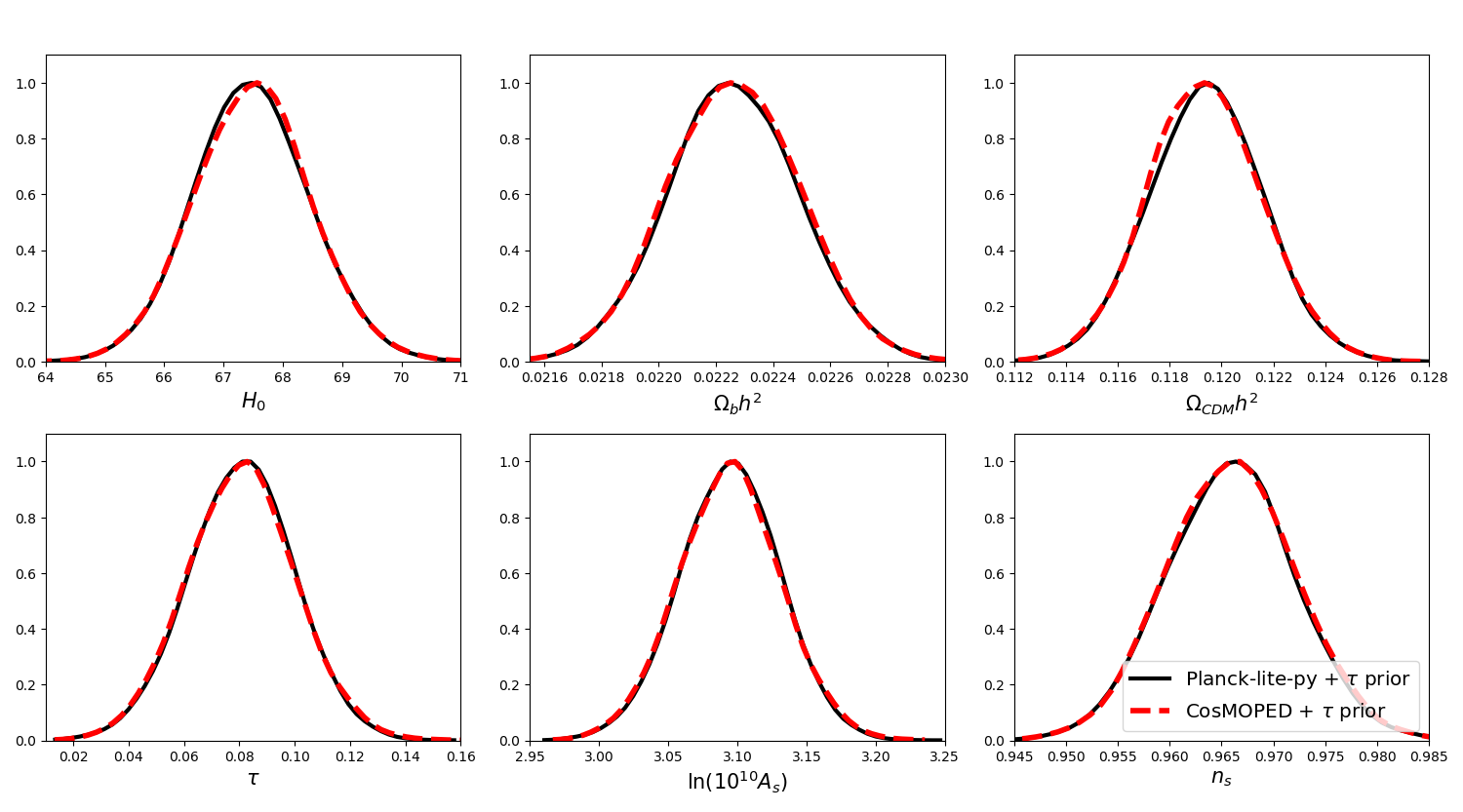}
  \caption{Posteriors on the \LCDM parameters from the compressed CosMOPED
  likelihood (red-dashed curves) and the Planck-lite-py Python implementation of
  the Plik\_lite likelihood with two low-$\ell$ bins (black solid curves). The
  parameter constraints agree to within 0.1$\sigma$, validating the CosMOPED
  compression scheme. A Gaussian prior of $\tau=0.067 \pm 0.023$ is included
  for both curves.}
   \label{fig:params_all_ell_cosmoped}
\end{figure*}

\section{Likelihood and parameters}
\label{sec:results}
We now describe the compressed likelihood and compare it to our Planck-lite-py
implementation of the \textit{Planck} 2015 Plik\_lite likelihood.
\subsection{Format of the likelihood}
Each compressed statistic $y_m$ is Gaussian distributed with unit variance.
The $y_m$'s are uncorrelated with each other by design, so the total likelihood
is the product of the likelihoods from each statistic. The likelihood of the
parameters given the (compressed) data thus takes a simple form
\be
-2 \ln \mathcal{L}= \sum_{m=1}^M \frac{(y_m-\langle y_m \rangle)^2}{2} +
\text{constant} ,
\ee
where $y_m$ is the compressed statistic from the data and $\langle y_m
\rangle=\boldsymbol{b_m}^t\boldsymbol{\mu}$  is the corresponding compressed
statistic from the model (the theoretical temperature power spectrum).

\subsection{Parameter constraints}
The parameter constraints for the six-parameter \LCDM model are shown in
Fig. \ref{fig:params_all_ell_cosmoped}. The compressed likelihood
(red-dashed curve) and the Python Plik\_lite implementation (black solid curve)
agree to within $0.1 \sigma$ for each parameter. Both likelihoods were sampled
with emcee, with a Gaussian prior on the optical depth $\tau= 0.067 \pm 0.023 $
and the low-$\ell$ bins described in Sec. \ref{sec:planck}.  The MOPED
compression that we have applied thus results in a likelihood that is
equivalent to the uncompressed case. As we showed earlier, the Python
Plik\_lite implementation is in good agreement with the full \textit{Planck}
temperature likelihood.

\subsection{Effect of fiducial model}
The theory vector $\boldsymbol\mu$ used to compute the compression vectors
depends on the fiducial model parameters used. If the fiducial model is wrong
then the compression is no longer optimal and the $M$ compressed statistics are
not exactly independent. However, in practice using a different fiducial model
does not have a significant effect on the compressed likelihood, in agreement
with the findings of \citet{zablocki_extreme_2016}. A shift of order 3$\sigma$
in the fiducial parameters has an insignificant effect on the conditional
probability slices obtained from the compressed likelihood.

\begin{figure}[!htbp]
  \centering
  \includegraphics[width=\linewidth]{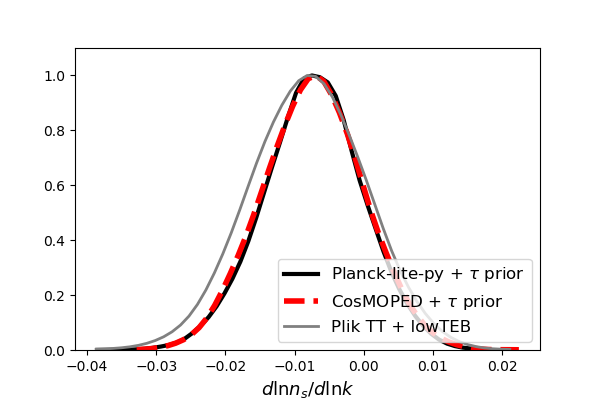}
  \caption{The constraints on running of the scalar spectral index from the
  compressed CosMOPED likelihood (red dashed) agree to within 0.1$\sigma$ with
  the uncompressed Planck-lite-py (black). A Gaussian prior of
  $\tau=0.067 \pm 0.023$ is used for CosMOPED and Planck-lite-py.
  The distribution using the public \textit{Planck} chains (TT+lowTEB, gray) is
  slightly broader due to the more accurate treatment of low-$\ell$ polarization.}
   \label{fig:running}
\end{figure}

\subsection{Non-\LCDM cases}
We demonstrate the application of this compression technique to a one parameter
extension to the \LCDM  case by sampling running of the scalar spectral index
$d\ln n_s/d \ln k$ in addition to the six \LCDM parameters shown above.

The results are shown in Fig. \ref{fig:running}, which compares the CosMOPED
constraints to the Planck-lite-py constraints (with the same $\tau$ prior as
above), as well as the \textit{Planck} TT+lowTEB chains from the 2015 data
release. The likelihoods show excellent agreement. The CosMOPED and
Planck-lite-py posteriors are slightly narrower than for the \textit{Planck}
TT+lowTEB  chains; this is because the Gaussian prior on $\tau$ which we use in
the CosMOPED likelihood comes from the low-$\ell$ likelihood assuming the
\LCDM model. For  \LCDM extension models with parameters that correlate with
$\tau$, this prior is slightly too narrow.

\section{Discussion and conclusion}
\label{sec:conclude}
We have demonstrated that the low-$\ell$ \textit{Planck} temperature data is
well represented by two Gaussian bins for simple cosmological models. We have
also shown that applying MOPED linear compression to the \textit{Planck} 2015
binned temperature power spectrum allows us to construct a simple likelihood
that depends on just one compression vector and one compressed statistic per
parameter of interest, and which is equivalent to the Plik-lite temperature
likelihood with two low-$\ell$ bins included. Because we do not directly
incorporate the low-$\ell$ polarization likelihood, we recommend including a
prior on the optical depth to reionization $\tau$ when using either of these
likelihoods.

We provide two public codes. The first is Planck-lite-py,  an implementation of
\textit{Planck}'s Plik\_lite likelihood in Python, with the option to include
the low-$\ell$ temperature data as two Gaussian bins. The second is CosMOPED,
which calculates the MOPED compression vectors for the CMB temperature power
spectrum and computes the compressed \textit{Planck} likelihood.

This method can easily be extended to incorporate the high-$\ell$ TE and EE
data which are also Gaussian distributed. It can also be used for the
\textit{Planck} 2018 likelihood, and the publicly available code will be
updated accordingly.

The MOPED data compression scheme provides a lightweight likelihood that can
easily be combined with other datasets. In addition, the compressed data can
be incorporated into a likelihood free inference framework which allows
parameters to be inferred based on forward simulations, without knowledge of
the form of the likelihood. In likelihood free inference it is useful to have
informative compressed statistics, because this makes the comparison of
simulations and data much less computationally intensive.

\section{Acknowledgments}
We thank Erminia Calabrese for useful comments. We acknowledge the use of 
data and code from the Planck Legacy Archive.

%

\end{document}